\documentclass[12pt,a4paper]{article}
\usepackage{amssymb,amsfonts,amsthm,amsmath}
\usepackage{textcomp}
\usepackage{setspace}

\usepackage{graphicx}

\def\hang{\hangindent\parindent}
 \def\rf{\par\noindent\hang}

\newtheorem{theorem}{Theorem}

\newtheorem{lemma}{Lemma}

\begin{document}

\baselineskip=18pt


\begin{center} \large{{\bf THE COVERAGE PROBABILITY OF CONFIDENCE INTERVALS IN ONE-WAY ANALYSIS OF COVARIANCE AFTER TWO F TESTS}}
\end{center}

\bigskip

\noindent Running Title:
CONFIDENCE INTERVALS AFTER TWO F TESTS

\bigskip

\begin{center}
\large \sc {Waruni Abeysekera, Paul Kabaila$^{*}$ and Oguzhan Yilmaz}
\end{center}

\begin{center}
{\it La Trobe University}
\end{center}

\vspace{12cm}


\noindent $^*$ Author to whom correspondence should be addressed.

\noindent Department of Mathematics and Statistics, La Trobe University, Victoria 3086, Australia.

\noindent e-mail: P.Kabaila@latrobe.edu.au

\noindent Facsimile: 3 9479 2466

\noindent Telephone: 3 9479 2594

\newpage

\begin{center}
\large{{\bf Summary}}
\end{center}

\noindent
Consider a one-way analysis of covariance model. Suppose that the parameter of interest $\theta$ is a specified linear
contrast of the expected responses, for a given value of the covariate. Also suppose that the inference
of interest is a $1-\alpha$ confidence interval for $\theta$.
The following two-stage
procedure has been proposed to determine the form of the model.
In Stage 1, we carry out an F test of the null hypothesis that the slopes are all zero
against the alternative hypothesis that they are not all zero.
If this null hypothesis is accepted then we assume that the slopes are all zero;
otherwise we proceed to Stage 2.
In Stage 2, we carry out an F test of the null hypothesis that the slopes
are all equal against the alternative hypothesis that
they are not all equal. If this null hypothesis is accepted then
we assume that the slopes
are all equal; otherwise this assumption is not made.
We present a general methodology for the examination of
the effect of this two-stage model selection procedure on the coverage probability of
a subsequently-constructed confidence interval for $\theta$, with nominal coverage $1-\alpha$.
This methodology is applied to a numerical example for which it is shown that this confidence interval is completely inadequate.

\bigskip


\newpage

\begin{center}
\large{{\bf 1. Introduction}}
\end{center}

Consider the one-way analysis of covariance model
\begin{equation}
\label{one_way_ancova}
Y_{ij}=a_{i} + b_{i} \left( x_{ij}-\bar x \right) + \varepsilon_{ij}
\end{equation}
\noindent where $Y_{ij}$ is the response of the $j^{th}$ experimental unit
$(j=1,...,n_{i})$ receiving treatment $i$ $(i=1,...,k)$, when the covariate takes the value $x_{ij}$.
The $\varepsilon_{ij}$ are independent and identically $N(0,\sigma^{2})$
distributed, where $\sigma^{2}$ is an unknown positive parameter. The $a_{i}$ and the slopes
$b_{i}$ are unknown parameters. Suppose that the parameter of interest $\theta$ is a specified linear
contrast of the expected responses, for a given value of the covariate. Also suppose that the inference
of interest is a $1-\alpha$ confidence interval (CI) for $\theta$.

Milliken \& Johnson (2002, Section 2.3) propose the following two-stage
procedure to determine the form of the model.
In Stage 1, we test the null hypothesis that the slopes $b_{i}$ are all zero
against the alternative hypothesis that they are not all zero. This test is carried out using an F statistic.
If this null hypothesis is accepted then we assume that the slopes $b_{i}$ are all zero;
otherwise we proceed to Stage 2.
In Stage 2, we test the null hypothesis that the slopes $b_{i}$
are all equal against the alternative hypothesis that
they are not all equal, which is also tested using an F statistic. If this null hypothesis is accepted then
we assume that the slopes $b_{i}$
are all equal; otherwise this assumption is not made.

Our aim is to examine the effect of this two-stage model selection procedure on the coverage probability (CP) of
a subsequently constructed CI for $\theta$, with nominal coverage $1-\alpha$.
This confidence interval is constructed on the assumption that the model selected by this two-stage
procedure had been given to us {\sl a priori} as the true model. This assumption is false and it may
lead to a CI with very poor coverage properties.

We present a general methodology for this examination in Sections 3 and 4.
Kabaila \& Farchione (2012) present a method (using numerical evaluation of multiple integrals) for evaluating
the CP of a CI for a scalar parameter constructed after a single preliminary
F test, in the context of a linear regression model.
This method does not extend to the present case of two preliminary F tests. We therefore need to
use Monte Carlo simulation to estimate the CP of the CI for $\theta$, with nominal coverage $1-\alpha$,
constructed after this two-stage model selection procedure.
In Section 3, we provide a simplified expression for the CP of this CI.
Let $\boldsymbol{\beta} = (a_1,\ldots,a_k,b_1,\ldots,b_k)$.
It follows from this simplified expression that this CP is a function of $\boldsymbol{\gamma} = \boldsymbol{\beta}/\sigma$.
Further, we show that this CP is a function of the parameter vector $(\gamma_{k+1},\ldots,\gamma_{2k})
=(b_1/\sigma, \ldots, b_k/\sigma)$.
In Section 4 we describe a new simulation method, using variance reduction by conditioning, for computing this CP.

In Section 2 this methodology is applied to an example (with number of treatments $k=3$)
which is used by Milliken \& Johnson (2002)
to illustrate their two-stage
procedure for determining the form of the model.
We suppose that the parameter of interest $\theta$ is the difference between the expected responses
of two subjects receiving the treatments 1 and 2, for the same specified value of the covariate.
We consider the CI for $\theta$, with nominal coverage $95\%$,
constructed after this two-stage procedure.
For both of the F tests used in this two-stage procedure, the significance level was chosen to be $10\%$.
The minimum CP of this CI is approximately 0.44, showing that it is completely inadequate.
Furthermore, as illustrated by Figures 1 and 2, the CP of this CI is far below 0.95 for a wide range
of centrally-located values of the parameter vector
$(\gamma_{4},\gamma_{5},\gamma_{6}) = (b_1/\sigma,\ldots,b_3/\sigma)$.

\medskip

\begin{center}
\large{{\bf 2. Numerical illustration for data taken from Milliken \& Johnson (2002)}}
\end{center}

The data provided by Milliken \& Johnson (2002) ``were generated to simulate real world applications
that we have encountered in our consulting experience''.
In this section, we consider data that is taken from Chapter 3 of Milliken \& Johnson (2002).
This data concerns the comparison of the
effectiveness of three exercise programs (treatments) on the heart rate of males
with ages in the range
from 28 to 35 years. A total of 24 males within this age range
were chosen and eight males were randomly assigned to each of the three treatments
labelled 1,2 and 3, so that $k=3$.
Since the aim was to compare exercise programs at a common initial resting heart rate, the initial heart rate of each of the subjects was
used as a covariate.

In their illustrative analysis of this data, Milliken \& Johnson (2002) begin with the one-way analysis of covariance
model \eqref{one_way_ancova} and perform the two-stage procedure (described in the introduction)
to determine the form of the model. We suppose that the parameter of interest
$\theta$ is the difference between the expected responses of two subjects receiving treatments 1 and 2,
for the same value $x^*$ of the covariate. We consider the CI for $\theta$, with nominal coverage $95\%$,
constructed after this two-stage procedure.
For both of the F tests performed in the two-stage procedure, the significance level was chosen to be $10\%$.

Let $Y_{1}^{*}$ and $Y_{2}^{*}$ denote the responses of two subjects receiving treatment 1 and 2, respectively,
for the same value $x^*$ of the covariate. That is
\begin{align*}
Y_{1}^{*}=a_{1} + b_{1} \left( x^*-\bar x \right) + \varepsilon_{1}^{*} \\
Y_{2}^{*}=a_{2} + b_{2} \left( x^*-\bar x \right) + \varepsilon_{2}^{*}
\end{align*}
\noindent where $\varepsilon_{1}^{*}$ and $\varepsilon_{2}^{*}$ are $N(0,\sigma^{2})$ distributed.
We define
\begin{equation*}
\theta=\text{E}(Y_{1}^{*})-\text{E}(Y_{2}^{*})=a_{1}-a_{2} + (b_{1}-b_{2}) \left( x^*-\bar x \right)
\end{equation*}
That is $\theta=\boldsymbol{a}^{\top}\boldsymbol{\beta}$, where
$\boldsymbol{a}= \big(1,-1,0,\left( x^*-\bar x \right),-\left( x^*-\bar x \right),0\big )$.
In this example we chose $\left( x^*-\bar x \right)$ such that $\mid x^*-\bar x \mid=$ maximum of all the $\mid x_{ij}^*-\bar x \mid$ values.

As we will show in Section 3 and Appendix C (and as already noted in the introduction) the CP of the CI constructed after
the two-stage procedure is a function of $(\gamma_4,\gamma_5,\gamma_6)$, where
$\boldsymbol{\gamma} = \boldsymbol{\beta}/\sigma$ ($\boldsymbol{\beta}=(a_1,a_2,a_3,b_1,b_2,b_3)$).
A search over the entire parameter space for the minimum CP of this CI is nearly impossible. In Appendix D,
we provide details of how we restrict the scope of this search, so that it becomes feasible. As shown in this
appendix, the minimum CP of this CI is achieved for $(\gamma_4,\gamma_5,\gamma_6) \in [-0.25,0.25]^3$.
Therefore, in the present section, we restrict our analysis of this CP function to $(\gamma_4,\gamma_5,\gamma_6) \in [-0.25,0.25]^3$.

We estimated the CPs for a grid of values of $(\gamma_4,\gamma_5,\gamma_6) \in [-0.25,0.25]^3$, using $M=10000$ simulation
runs for each paramater value.
When these estimated CPs are plotted using a 3-D Scatter plot, it is observed that the minimum CP is approximately $0.44$
and that the CP is small for values of $(\gamma_4,\gamma_5,\gamma_6)$ that lie close to two parallel straight lines in the 3-D space of parameters.
The equations of these two lines were found by fitting linear regression lines to the parameter values that gave
estimated CPs less than $0.6$. The fitted equations for the two lines were found to be as follows.
\begin{align*}
& \bold{Line \ 1}: \ \gamma_{4}=c, \ \gamma_{5}=0.088+c, \ \gamma_{6}=0.041+c \\
& \bold{Line \ 2}: \ \gamma_{4}=c, \ \gamma_{5}=-0.088+c, \ \gamma_{6}=-0.041+c
\end{align*}
where $-0.25 \leq c \leq 0.25$. Then the CPs were re-estimated using $M=10000$ simulation runs
for each of a grid of parameter values on
these two lines, and plotted in Figures 1 and 2. Figure 1 is a plot of the estimated CP for
parameters $(\gamma_4,\gamma_5,\gamma_6)$ on Line 1. Figure 2 is a plot of the estimated CP for
parameters $(\gamma_4,\gamma_5,\gamma_6)$ on Line 2. From Figure 1, the minimum CP on Line 1 is estimated to occur at $c=-0.0453$
i.e. at $(\gamma_4,\gamma_5,\gamma_6)=(-0.0453,0.0427,-0.0043)$. This minimum was estimated to be $0.4384$, with standard error $0.0035$.
From Figure 2, the minimum CP on Line 2 is estimated to occur at $c=0.0468$ i.e. at $(\gamma_4,\gamma_5,\gamma_6)=(0.0468,-0.0412,0.0058)$.
This minimum was estimated to be $0.4385$, with standard error $0.0035$. Thus the minimum CP is, to a good approximation, $0.4385$.
In other words, the CI for $\theta$ have minimum CP far below 0.95, showing that it is completely
inadequate. Furthermore, as illustrated by Figures 1 and 2, the CP of this CI is far below 0.95 for a wide range
of centrally-located values of the parameter vector $(\gamma_{4},\gamma_{5},\gamma_{6})$.
The fact that, for this example, the lowest CPs lie close to two parallel straight lines is investigated further in Appendix A.

\newpage

\begin{figure}[h!]
\begin{center}$
\begin{array}{cc}
\includegraphics[scale=0.35]{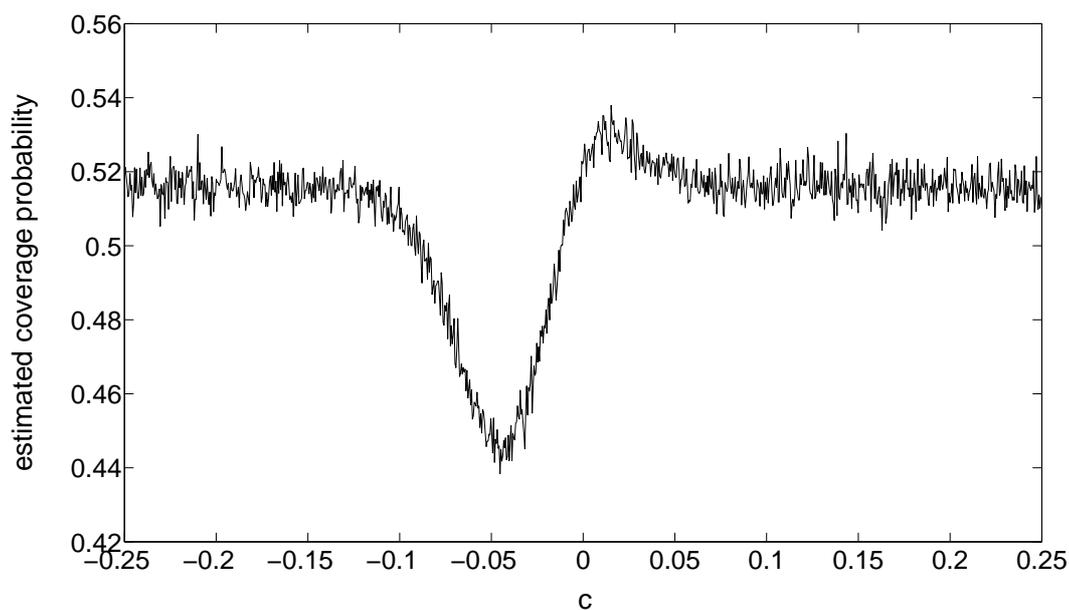}
\end{array}$
\end{center}
\caption{Plot of the estimated  coverage probability for the parameter vector
$(\gamma_{4},\gamma_{5},\gamma_{6})$ on $\bold{Line \ 1}: \ \gamma_{4}=c, \ \gamma_{5}=0.088+c, \ \gamma_{6}=0.041+c$ }
\end{figure}
\begin{figure}[h!]
\begin{center}$
\begin{array}{cc}
\includegraphics[scale=0.35]{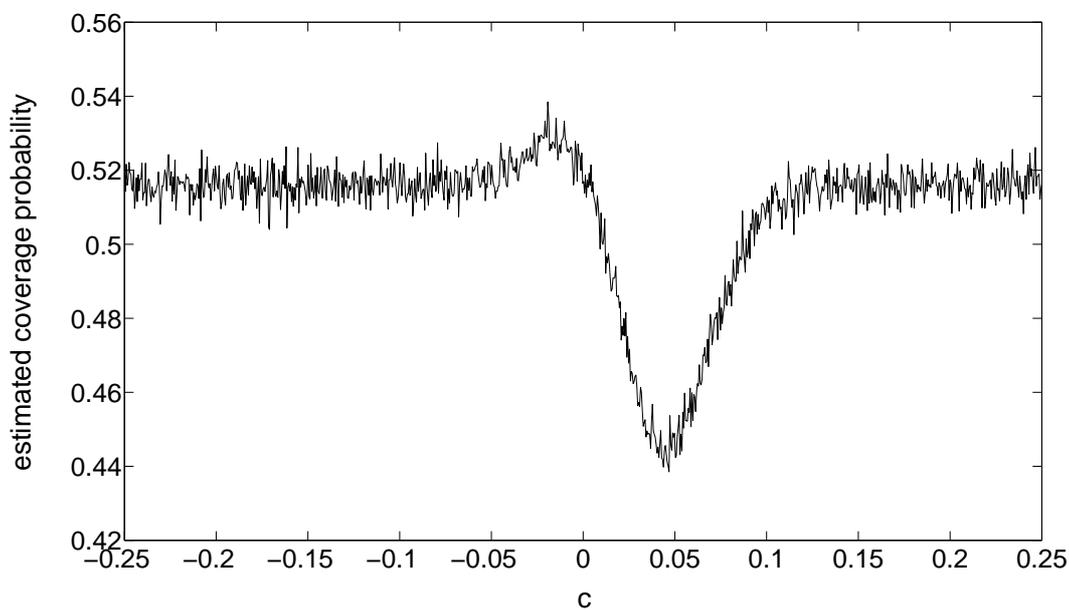}
\end{array}$
\end{center}
\caption{Plot of the estimated coverage probability for the parameter vector $(\gamma_{4},\gamma_{5},\gamma_{6})$ on $\bold{Line \ 2}: \ \gamma_{4}=c, \ \gamma_{5}=-0.088+c, \ \gamma_{6}=-0.041+c$ }
\end{figure}

\newpage

\bigskip

\begin{center}
\large{{\bf 3. Simplified expression for the coverage probability of the confidence interval for $\theta$}}
\end{center}

In this section we consider the general situation described in the introduction.
The CI for $\theta$ has three different forms, depending on the model resulting from the two-stage
procedure. To find an expression for the coverage probability of this confidence interval, we use
the law of total probability (cf Section 2 of Giri \& Kabaila, 2008). This coverage probability is a function of the $2k+1$
dimensional parameter vector $(\boldsymbol{\beta}, \sigma^2)$. By dividing by $\sigma$ in the appropriate
way, we show that this coverage probability is, in fact, a function of $\boldsymbol{\gamma}=\boldsymbol{\beta}/\sigma$.
Then we prove Theorem 1, which states that this CP is a function of the parameter vector
$(\gamma_{k+1},\ldots,\gamma_{2k}) = (b_1/\sigma,\ldots,b_k/\sigma)$.
This reduces the dimensionality of the parameter space over which we search for the minimum CP.

For the model (1) that we consider, there are $k$ treatments with $n_{i}$ experimental units allocated to
the $i^{th}$ treatment, so that the total number of measurements of the response is $n=\displaystyle \sum_{i=1}^k n_{i}$.
We express the model as $\bold Y=\bold X \boldsymbol\beta + \boldsymbol\varepsilon$, where $\bold Y=\left(Y_{11},\ldots,Y_{1n_{1}},\ldots,Y_{k1},\ldots,Y_{kn_{k}}\right)$,
$\bold X$ is an $n \times 2k$ design matrix,
$\boldsymbol{\beta}= (\beta_1, \ldots, \beta_{2k}) = \left(a_{1},\ldots,a_{k},b_{1},\ldots,b_{k}\right)$, and
$\boldsymbol\varepsilon=\left(\varepsilon_{11},\ldots,\varepsilon_{1n_{1}},\ldots,\varepsilon_{k1},\ldots,\varepsilon_{kn_{k}}\right)$.
Let $\boldsymbol{\hat\beta}$ denote the least squares estimator of $\boldsymbol{\beta}$.
Also let $\hat\Sigma^2=(\bold{Y}-\boldsymbol{X\hat\beta})^{\top}(\bold{Y}-\boldsymbol{X\hat\beta})/m$, where $m=n-2k$.

The preliminary F tests are taken to follow the two-stage procedure described in the introduction.
Accordingly, the null hypothesis in Stage 1 is $H_{0\tau}: b_{1}=b_{2}=\cdots=b_{k}=0$.
Let $\boldsymbol\tau=\boldsymbol{C_{\tau}}^{\top}\boldsymbol{\beta}$. Here, $\boldsymbol{C_{\tau}}= \left[ \ \bold{0} \ \text{\textbrokenbar} \ \boldsymbol{\textbf{I}_{k}} \ \right]$,
where $\bold{0}$ is the $k \times k$ zero matrix and
$\boldsymbol{\textbf{I}_{k}}$ is the $k \times k$ identity matrix.
In other words, $\boldsymbol{C_{\tau}}$ is the $k \times 2k$ matrix defined such that
$\boldsymbol{C_{\tau}}^{\top}\boldsymbol{\beta}= (b_{1},\ldots,b_{k})= (\beta_{k+1},\ldots,\beta_{2k})$.
Thus the test can be re-expressed as $H_{0\tau}: \boldsymbol{\tau}=\boldsymbol{0}$ against
$H_{1\tau}: \boldsymbol{\tau} \not= \boldsymbol{0}$. The F test for testing this hypothesis has the following test statistic
\begin{equation*}
F_{\tau}= \left( \frac{m}{k} \right) \frac{\left( \boldsymbol{\hat\tau} / \sigma \right)^{\top} \boldsymbol{V_{22}}^{-1} \left( \boldsymbol{\hat\tau} / \sigma \right)}{\left( m \hat\Sigma^2 / \sigma^2 \right)}
\end{equation*}
where $\boldsymbol{\hat\tau}=\boldsymbol{C_{\tau}}^{\top} \boldsymbol{\hat\beta}$, and
$\boldsymbol{V_{22}}=(1/\sigma^2)\text{Cov}(\boldsymbol{\hat\tau})=\boldsymbol{C_{\tau}}^{\top}(\boldsymbol{X}^{\top}\boldsymbol{X})^{-1}\boldsymbol{C_{\tau}}$.
The null hypothesis $H_{0\tau}$ is rejected if $F_{\tau}>\ell_{\tau}$ (accepted otherwise).
%

The null hypothesis in Stage 2 is $H_{0\xi}: b_{1}=b_{2}=\cdots=b_{k}$.
We let $\boldsymbol\xi=\boldsymbol{C_{\xi}}^{\top}\boldsymbol{\beta}$. Here
$\boldsymbol{C_{\xi}}=\left[ \ \bold{0} \ \text{\textbrokenbar} \ \bold{1}  \ \text{\textbrokenbar} \ -\boldsymbol{\textbf{I}_{(k-1)}} \ \right]$,
where $\bold{0}$ is the $(k-1) \times k$ zero matrix, $\bold{1}$ is the $(k-1)$ vector of 1's  and
$\boldsymbol{\textbf{I}_{k-1}}$ is the $(k-1) \times (k-1)$ identity matrix.
In other words, $\boldsymbol{C_{\xi}}$ is the $(k-1) \times 2k$ matrix defined such that $\boldsymbol{C_{\xi}}^{\top}\boldsymbol{\beta}=(b_{1}-b_{2},\ldots,b_{1}-b_{k})=(\beta_{k+1}-\beta_{k+2},\ldots,\beta_{k+1}-\beta_{2k})$.
Hence the test is re-expressed as $H_{0\xi}: \boldsymbol{\xi}=\boldsymbol{0}$ against
$H_{1\xi}: \boldsymbol{\xi} \not= \boldsymbol{0}$. The F test
for testing this hypothesis has the following test statistic
\begin{equation*}
F_{\xi}= \left( \frac{m}{k-1} \right) \frac{\left( \boldsymbol{\hat\xi} / \sigma \right)^{\top} \boldsymbol{W_{22}}^{-1} \left( \boldsymbol{\hat\xi} / \sigma \right)}{\left( m \hat\Sigma^2 / \sigma^2 \right)}
\end{equation*}
\noindent where $\boldsymbol{\hat\xi}=\boldsymbol{C_{\xi}}^{\top} \boldsymbol{\hat\beta}$, and
$\boldsymbol{W_{22}}=(1/\sigma^2)\text{Cov}(\boldsymbol{\hat\xi})=\boldsymbol{C_{\xi}}^{\top}(\boldsymbol{X}^{\top}\boldsymbol{X})^{-1}\boldsymbol{C_{\xi}}$.
The null hypothesis $H_{0\xi}$ is rejected if $F_{\xi}>\ell_{\xi}$ (accepted otherwise).

The following three events form a partition of the sample space $\Omega$, induced by this two-stage
procedure:
\begin{align*}
&A=\{ \omega\in\Omega : F_{\tau}(\omega)\leq\ell_{\tau} \} \\
&B=\{ \omega\in\Omega : F_{\tau}(\omega)>\ell_{\tau}, F_{\xi}(\omega)\leq\ell_{\xi} \} \\
&C=\{ \omega\in\Omega : F_{\tau}(\omega)>\ell_{\tau}, F_{\xi}(\omega)>\ell_{\xi} \}
\end{align*}

\noindent The parameter of interest is the linear contrast $\theta=\boldsymbol{a}^{\top}\boldsymbol{\beta}$.
Let $\hat{\Theta} = \boldsymbol{a}^{\top} \boldsymbol{\hat\beta}$.
Let $v_{11}=w_{11}=(1/\sigma^2)\text{Var}(\hat{\Theta})=\boldsymbol{a}^{\top}(\boldsymbol{X}^{\top}\boldsymbol{X})^{-1}\boldsymbol{a}$,
$\boldsymbol{v_{21}}=(1/\sigma^2)\text{E}\big((\boldsymbol{\hat\tau}-\boldsymbol{\tau})(\hat\Theta-\theta)\big)
=\boldsymbol{C_{\tau}}^{\top}(\boldsymbol{X}^{\top}\boldsymbol{X})^{-1}\boldsymbol{a}$
and
$\boldsymbol{w_{21}}=(1/\sigma^2)\text{E}\big((\boldsymbol{\hat\xi}-\boldsymbol{\xi})(\hat\Theta-\theta)\big)
=\boldsymbol{C_{\xi}}^{\top}(\boldsymbol{X}^{\top}\boldsymbol{X})^{-1}\boldsymbol{a}$ .
Also let $\boldsymbol{\hat\beta_{\tau}}$ denote the value of $\boldsymbol{\beta}$ that minimizes
$R(\boldsymbol{\beta})=(\bold{Y}-\boldsymbol{X\beta})^{\top}(\bold{Y}-\boldsymbol{X\beta})$
when $\boldsymbol{\tau} = \boldsymbol{0}$.  As is well-known, $\boldsymbol{\hat\beta_{\tau}}= \boldsymbol{G_{\tau}} \hat{\boldsymbol{\beta}}$, where
$\boldsymbol{G_{\tau}}=\boldsymbol{I}-(\boldsymbol{X}^{\top}\boldsymbol{X})^{-1} \boldsymbol{C_{\tau}}\boldsymbol{V_{22}}^{-1}\boldsymbol{C_{\tau}}^{\top}$. When
$\boldsymbol{\tau} = \boldsymbol{0}$, the standard $1-\alpha$ confidence interval for $\theta$ is
\begin{equation*}
I_{\tau} = \left [ \boldsymbol{a}^{\top} \boldsymbol{\hat\beta_{\tau}} \pm t(m+k)
\sqrt{\frac{R(\boldsymbol{\hat\beta_{\tau}})}{m+k}} \sqrt{v^{*}} \right ],
\end{equation*}
where $v^{*}=v_{11}-\boldsymbol{v_{21}}^{\top}\boldsymbol{V_{22}}^{-1}\boldsymbol{v_{21}}$ and $t(r)$ is the quantile defined by
$\text{Pr}\left( T \leq t(r) \right)=1-\alpha/2$ for $T \sim t_r$ .
Similarly, let $\boldsymbol{\hat\beta_{\xi}}$ denote the value of $\boldsymbol{\beta}$ that minimizes
$R(\boldsymbol{\beta})$
when $\boldsymbol{\xi} = \boldsymbol{0}$.  As is well-known, $\boldsymbol{\hat\beta_{\xi}}= \boldsymbol{G_{\xi}} \hat{\boldsymbol{\beta}}$, where
$\boldsymbol{G_{\xi}}=\boldsymbol{I}-(\boldsymbol{X}^{\top}\boldsymbol{X})^{-1} \boldsymbol{C_{\xi}}\boldsymbol{W_{22}}^{-1}\boldsymbol{C_{\xi}}^{\top}$.
When $\boldsymbol{\xi} = \boldsymbol{0}$, the standard $1-\alpha$ confidence interval for $\theta$ is
\begin{equation*}
I_{\xi} = \left [ \boldsymbol{a}^{\top} \boldsymbol{\hat\beta_{\xi}} \pm t(m+k-1)
\sqrt{\frac{R(\boldsymbol{\hat\beta_{\xi}})}{m+k-1}} \sqrt{w^{*}} \right ],
\end{equation*}
where $w^{*}=w_{11}-\boldsymbol{w_{21}}^{\top}\boldsymbol{W_{22}}^{-1}\boldsymbol{w_{21}}$. The standard $1-\alpha$ confidence interval for $\theta$, when fitting
the full model to the data is
\begin{equation*}
I = \left [ \boldsymbol{a}^{\top} \boldsymbol{\hat\beta} \pm t(m) \sqrt{v_{11}} \hat{\Sigma}
\right ].
\end{equation*}

The CI for $\theta$, with nominal coverage $1-\alpha$, constructed after the two-stage model selection
procedure is given by the following expression.
\begin{equation*}
CI(\omega) =
\begin{cases}
I_{\tau}(\omega) \ \ \text{if} \ \omega\in A \\
I_{\xi}(\omega) \ \ \text{if} \ \omega\in B \\
I(\omega) \ \ \ \text{if} \ \omega\in C
\end{cases}
\end{equation*}
Therefore, using the law of total probability the CP of this CI can be expressed as
\begin{equation}
\label{law_tot_prob}
\text{Pr}\left( \theta \in \text{CI} \right)
=\text{Pr}\left( \theta \in I_{\tau}, A \right)+\text{Pr}\left( \theta \in I_{\xi}, B \right)+\text{Pr}\left( \theta \in I, C \right)
\end{equation}
A simplified expression for this CP is obtained by substituting the simplified expressions, presented and derived in Appendix B, for the events
$\{\theta \in I_{\tau}\}$, $\{\theta \in I_{\xi}\}$ and $\{\theta \in I\}$.
This simplified expression implies that this CP is a function of $\boldsymbol{\gamma} = \boldsymbol{\beta}/\sigma$.

\begin{theorem}

 The CP of the CI resulting from the two-stage procedure is (for given design matrix $\boldsymbol{X}$)
 a function of
 the parameter vector $(\gamma_{k+1},\ldots,\gamma_{2k})$.

\end{theorem}

\noindent The proof of this theorem is provided in Appendix C.

\bigskip

\begin{center}
\large{{\bf 4. New simulation method for estimating the CP of the CI for $\boldsymbol{\theta}$}}
\end{center}

In this section we describe a new simulation method for estimating the CP of the CI for $\theta$, with nominal coverage $1-\alpha$,
constructed after the two-stage model selection procedure. This new method uses variance reduction by conditioning.
Variance reduction by conditioning is described, for example, on p.629 of Ross (2000).
\medskip

Let $\boldsymbol{Q}=\boldsymbol{\hat\tau}/\sigma$ and $D=m \hat\Sigma^2 / \sigma^2$. In Appendix E, we provide expressions for
the following conditional probabilities:
\begin{align*}
& p_{\tau}(\boldsymbol{q},d)=\text{Pr} \left( \theta \in I_{\tau}, A \ \vert \ \boldsymbol{Q}=\boldsymbol{q}, \ D=d \right) \\
& p_{\xi}(\boldsymbol{q},d)=\text{Pr} \left( \theta \in I_{\xi}, B \ \vert \ \boldsymbol{Q}=\boldsymbol{q}, \ D=d \right) \\
& p(\boldsymbol{q},d)=\text{Pr} \left( \theta \in I, C \ \vert \ \boldsymbol{Q}=\boldsymbol{q}, \ D=d \right).
\end{align*}
The CP of the CI for $\theta$ is
\begin{equation}
\label{comp_cov_pr}
\text{E}\left( p_{\tau}(\boldsymbol{Q},D) \right)+\text{E}\left( p_{\xi}(\boldsymbol{Q},D) \right)+\text{E}\left( p(\boldsymbol{Q},D) \right).
\end{equation}
We could estimate this CP by adding simulation estimates of each of the terms making up this sum. Obviously, \eqref{comp_cov_pr} is equal to
$\text{E}\big( p_{\tau}(\boldsymbol{Q},D) +  p_{\xi}(\boldsymbol{Q},D) +  p(\boldsymbol{Q},D) \big)$. Thus, an alternative simulation estimate
of this CP is the sample average of $M$ independent observations of $p_{\tau}(\boldsymbol{Q},D) +  p_{\xi}(\boldsymbol{Q},D) +  p(\boldsymbol{Q},D)$.
In the context of the example described in Section 2, this is the more efficient simulation method.


\newpage

\begin{center}
\large{{\bf 5. Discussion}}
\end{center}

The literature on the effect of preliminary model selection (using, for example, hypothesis tests or minimizing a criterion such as AIC)
on CIs is reviewed by Kabaila (2009). It is commonly the case that preliminary model selection has a detrimental
effect on the CP of these CIs. However, each case (specified by a model, a model selection procedure
and a parameter of interest)
needs to be considered individually on its merits.

In the present paper, we consider the two-stage model selection procedure proposed by Milliken \& Johnson (2002, Section 2.3)
in the context of a one-way analysis of covariance model. This procedure involves the use of two F tests. We present a general
methodology for examining the effect of this procedure on the CP of a subsequently-constructed CI for a specified linear
contrast of the expected responses, for a given value of the covariate. This general methodology has the following two components.
The first component is a theorem that states that this CP is a function of a $k$-dimensional parameter
vector, rather that a $(2k+1)$-dimensional
parameter vector (as one might initially suppose), where $k$ is the number of treatments.
This increases the feasibility of examining the coverage probability function
closely, including (a) finding its minimum and (b) finding those parts of the parameter space where it is far below nominal.
The second component is a new simulation method, using variance reduction by conditioning, for computing this CP. Although the
derivation of this simulation method is complicated, it brings important benefits in the form of increased simulation efficiency.
This general methodology extends in the obvious way to any two-stage model selection procedure that uses two F tests
(in a similar way to that described in the introduction)
in the context
of any linear regression model with
independent and identically normally distributed random errors.

We have applied this general methodology to data taken from Chapter 3 of Milliken \& Johnson (2002),
where the difference of expected responses for treatments 1 and 2 is, for a given value of the covariate,
specified as the parameter of interest.
We have shown that the CP of the CI for this contrast is far below nominal for a wide range of centrally-located parameter
values. This throws doubt on the utility of the two-stage model selection procedure proposed by Milliken \& Johnson (2002, Section 2.3).


\newpage

\begin{center}
\large{{\bf Appendix A: Influence of $\boldsymbol{a}$ on the locus of values of $\boldsymbol{(\gamma_4,\gamma_5,\gamma_6)}$
for which the CP is small}}
\end{center}

The locus of values of $(\gamma_4,\gamma_5,\gamma_6)$, for which the CP is small, was found to depend
mainly on the linear contrast $\theta$ under consideration. We see this by giving different values to $\boldsymbol{a}$,
the vector of contrast coefficients. For example, when $\boldsymbol{a}=(1,0,0,\left( x^*-\bar x \right),0,0)$,
the lower estimated CPs are found lie close to two parallel planes that are parallel to the plane that includes the
$\gamma_{5}$ and $\gamma_{6}$ axes.
Also, when $\boldsymbol{a}=(0,1,0,0,\left( x^*-\bar x \right),0)$, the lower estimated CPs lie close to two parallel planes that are parallel
to the plane that includes the
$\gamma_{4}$ and $\gamma_{6}$ axes.

\bigskip

\begin{center}
\large{{\bf Appendix B: Simplified expressions for the events
$\boldsymbol{\{\theta \in I_{\tau}\}}$, $\boldsymbol{\{\theta \in I_{\xi}\}}$ and $\boldsymbol{\{\theta \in I\}}$}}
\end{center}

As in Sections 3 and 4,
let $\boldsymbol{\hat\gamma}=\boldsymbol{\hat\beta}/\sigma$, $\bold{Q}=\boldsymbol{\hat\tau}/\sigma$ and $D=m \hat\Sigma^2 / \sigma^2$.
We obtain a simplified expression for the event $\{\theta \in I_{\tau}\}$ as follows.
It follows from $\boldsymbol{\hat\beta_{\tau}}=\boldsymbol{G_{\tau}\hat\beta}$ that
$\boldsymbol{a}^{\top} \boldsymbol{\hat\beta_{\tau}}/\sigma = \boldsymbol{a}^{\top}\boldsymbol{G_{\tau}\hat\gamma}$.
Since
$R(\boldsymbol{\hat\beta_{\tau}})
= R(\boldsymbol{\hat\beta})+(\boldsymbol{\hat\beta}-\boldsymbol{\hat\beta_{\tau}})^{\top}\boldsymbol{X}^{\top}\boldsymbol{X}(\boldsymbol{\hat\beta}-\boldsymbol{\hat\beta_{\tau}})
= m \hat\Sigma^2 + \boldsymbol{\hat\tau}^{\top} \boldsymbol{V_{22}}^{-1} \boldsymbol{\hat\tau}$,
\begin{equation*}
R(\boldsymbol{\hat\beta_{\tau}}) / \sigma^2 = D + \boldsymbol{Q}^{\top} \boldsymbol{V_{22}}^{-1} \boldsymbol{Q}.
\end{equation*}
Thus
\begin{align*}
& \{\theta \in I_{\tau}\} = \{\theta/\sigma \in I_{\tau}/\sigma\} \\
& \ \ \ \ \ \ \ \ \ \ \ = \left\{ \boldsymbol{a}^{\top}\boldsymbol{\beta}/\sigma \in \left[ \boldsymbol{a}^{\top} \boldsymbol{\hat\beta_{\tau}}/\sigma \pm t(m+k)
\sqrt{\frac{R(\boldsymbol{\hat\beta_{\tau}})/\sigma^2}{m+k}} \sqrt{v^{*}} \right] \right\} \\
& \ \ \ \ \ \ \ \ \ \ \ = \left\{ \boldsymbol{a}^{\top}\boldsymbol{\gamma} \in \left[ \boldsymbol{a}^{\top}\boldsymbol{G_{\tau}\hat\gamma} \pm t(m+k)
\sqrt{\frac{D + \boldsymbol{Q}^{\top} \boldsymbol{V_{22}^{-1}} \boldsymbol{Q}}{m+k}} \sqrt{v^{*}} \right] \right\}.
\end{align*}
The following simplified expressions for the events
$\{\theta \in I_{\xi}\}$ and $\{\theta \in I\}$ can be obtained using a similar method. Let
$\boldsymbol{U}= \left[ \ \bold{1} \ \text{\textbrokenbar} \ -\boldsymbol{\textbf{I}_{k-1}} \ \right]$, where
$\bold{1}$ is the $(k-1)$ vector of 1 s
and $\boldsymbol{\textbf{I}_{k-1}}$ is the $(k-1) \times (k-1)$ identity matrix. Therefore $\boldsymbol{\xi}=\boldsymbol{U}\boldsymbol{\tau}$ and hence
$\boldsymbol{\hat\xi}/\sigma=\boldsymbol{U}\boldsymbol{Q}$.
\begin{align}
\label{event_theta_in_I_ksi}
\{\theta \in I_{\xi}\} &=
\left\{ \boldsymbol{a}^{\top}\boldsymbol{\gamma} \in \left[ \boldsymbol{a}^{\top}\boldsymbol{G_{\xi}\hat\gamma} \ \pm \
t(m+k-1)\sqrt{\frac{D + \boldsymbol{Q}^{\top} \boldsymbol{U}^{\top} \boldsymbol{W_{22}}^{-1} \boldsymbol{U}\boldsymbol{Q}}{m+k-1}}
\sqrt{w^{*}} \right] \right\} \\
\label{event_theta_in_I}
\{\theta \in I\} &=
\left\{ \boldsymbol{a}^{\top}\boldsymbol{\gamma} \in \left[ \boldsymbol{a}^{\top}\boldsymbol{\hat\gamma} \ \pm \
t(m) \sqrt{\frac{D}{m}} \sqrt{v_{11}} \right] \right\}
\end{align}

\bigskip

\begin{center}
\large{{\bf Appendix C: Proof of Theorem 1}}
\end{center}

It follows from \eqref{law_tot_prob} and the simplified expressions for the events
$\{\theta \in I_{\tau}\}$, $\{\theta \in I_{\xi}\}$ and $\{\theta \in I\}$ given in Appendix B
that the CP of the CI resulting from the two-stage model selection procedure is a
function of $\boldsymbol{\gamma}$.
In the present appendix, we prove that $\text{Pr}\left( \theta \in I, C \right)$ is a function of $(\gamma_{k+1},\ldots,\gamma_{2k})$.
In the same manner, it can be proved that $\text{Pr}\left( \theta \in I_{\tau}, A \right)$ and
$\text{Pr}\left( \theta \in I_{\xi}, B \right)$ are also
functions of $(\gamma_{k+1},\ldots,\gamma_{2k})$. It follows from \eqref{law_tot_prob} that
the CP is also a function of $(\gamma_{k+1},\ldots,\gamma_{2k})$.

The occurrence or otherwise of the event $C=\{F_{\tau}>\ell_{\tau}, F_{\xi}>\ell_{\xi} \}$ is determined by the statistics
$F_{\tau}$ and $F_{\xi}$, defined in Section 3.
Note that $F_{\tau}$ is a function of $\boldsymbol{\hat\tau}/\sigma$ and $m \hat\Sigma^2 / \sigma^2$ and
$F_{\xi}$ is a function of $\boldsymbol{\hat\xi}/\sigma$ and $m \hat\Sigma^2 / \sigma^2$.
As in Appendix B, $\boldsymbol{Q}=\boldsymbol{\hat\tau}/\sigma=(\hat\gamma_{k+1},\ldots,\hat\gamma_{2k})$,
$\boldsymbol{\hat\xi}/\sigma=\boldsymbol{U}\boldsymbol{Q}=(\hat\gamma_{k+1}-\hat\gamma_{k+2},\ldots,\hat\gamma_{k+1}-\hat\gamma_{2k})$
and $D=m \hat\Sigma^2 / \sigma^2$.
Therefore, $F_{\tau}$ and $F_{\xi}$ are functions of $(\boldsymbol{Q},D)$.
Thus, occurrence or otherwise of the event $C$ is determined by the random quantities
\begin{equation*}
(\hat\gamma_{k+1} \ , \ \ldots \ , \ \hat\gamma_{2k})  \ \text{and} \ D.
\end{equation*}
In other words, the occurrence or otherwise of the event $C$ is determined by the quantities
\begin{equation*}
(\gamma_{k+1},\ldots,\gamma_{2k}),
\big((\gamma_{k+1}-\hat\gamma_{k+1}), \ldots , (\gamma_{2k}-\hat\gamma_{2k}) \big)  \ \text{and} \ D.
\end{equation*}
It follows from \eqref{event_theta_in_I} that the occurrence or otherwise of the event $\{\theta \in I\}$
is determined by the random quantities
\begin{equation*}
\boldsymbol{\gamma-\hat\gamma} \ \text{and} \ D.
\end{equation*}
Therefore, the occurrence or otherwise of the event $\{\theta \in I\} \cap C$ is determined by the quantities
\begin{equation*}
(\gamma_{k+1},\ldots,\gamma_{2k}),
\boldsymbol{\gamma-\hat\gamma}  \ \text{and} \ D.
\end{equation*}
Since $\boldsymbol{\gamma-\hat\gamma}$ and $D$ are independent random vectors with
\begin{equation*}
\boldsymbol{\gamma}-\hat{\boldsymbol{\gamma}}
= (\boldsymbol{\beta}-\hat{\boldsymbol{\beta}})/\sigma \sim N(\boldsymbol{0} , (\boldsymbol{X}^{\top}\boldsymbol{X})^{-1} )
\end{equation*}
and $D$ has a $\chi_m^2$ distribution, $\text{Pr}\left(\theta \in I, C \right)$ is (for given
design matrix $\boldsymbol{X}$)
a function of $(\gamma_{k+1},\ldots,\gamma_{2k})$.

\bigskip

\begin{center}
\large{{\bf Appendix D: Search for the minimum coverage probability}}
\end{center}

The CP of the CI described in Section 2 is a function of $(\gamma_{4},\gamma_{5},\gamma_{6})$. It is very
difficult (if not impossible) to carry out a computational search for the minimum CP of this CI over
$(\gamma_{4},\gamma_{5},\gamma_{6}) \in \mathbb{R}^3$. To carry out this search, we need to restrict
the scope of this search. The two purposes of this appendix are to (a) describe the method used to
restrict this search and (b) report the result of carrying out this restricted search for the
minimum CP.

Our restriction of this search is based on the following simple result.

\begin{lemma}

For any events $S$ and $T$, $0 \le \text{Pr}(S) - \text{Pr}(S \cap T) \le \text{Pr}(T^c)$.
Consequently, if $\text{Pr}(T)$ is close
to 1 then $\text{Pr}(S \cap T)$ is close to $\text{Pr}(S)$.

\end{lemma}

\noindent According to \eqref{law_tot_prob}, the coverage probability $\text{Pr}(\theta \in CI)$ is equal to
\begin{equation}
\label{CP}
\text{Pr}\big(\theta \in I_{\tau}, \, F_{\tau} \le \ell_{\tau} \big)
+\text{Pr}\big(\theta \in I_{\xi}, \,  F_{\tau} > \ell_{\tau}, \, F_{\xi} \le \ell_{\xi} \big)
+\text{Pr}\big(\theta \in I, \,  F_{\tau} > \ell_{\tau}, \, F_{\xi} > \ell_{\xi} \big).
\end{equation}
It can be shown that $\text{Pr}\big(F_{\tau} \le \ell_{\tau} \big) \approx 1$, for any value of $(\gamma_{4},\gamma_{5},\gamma_{6})$
outside the cube $[-0.25, 0.25]^3$. It follows from \eqref{CP} and Lemma 1 that
\begin{equation}
\label{CP_outside_cube}
\text{Pr}(\theta \in CI) \approx \text{Pr}\big(\theta \in I_{\xi}, \, F_{\xi} \le \ell_{\xi} \big)
+\text{Pr}\big(\theta \in I, \, F_{\xi} > \ell_{\xi} \big),
\end{equation}
for any value of $(\gamma_{4},\gamma_{5},\gamma_{6})$ outside this cube. In other words, for any value of
$(\gamma_{4},\gamma_{5},\gamma_{6})$ outside this cube, the computation of the minimum CP can, to a very
good approximation, be based on the assumption that the model selection procedure consists only of the second F test.

We now use the following result.

\begin{lemma}

If the model selection procedure consists only of the second F test then the CP of the subsequently constructed CI
for $\theta$ is a function of $(\gamma_5 - \gamma_4, \gamma_6 - \gamma_4)$.

\end{lemma}

\noindent For the sake of brevity, we omit the proof of this result.
It can be shown that $\text{Pr}\big(F_{\xi} > \ell_{\xi} \big) \approx 1$ for any value of
$(\gamma_5 - \gamma_4, \gamma_6 - \gamma_4)$ outside the square $[-0.2,0.2]^2$.
By Lemma 1, if the model selection procedure consists only of the second F test
then the CP of the subsequently constructed CI for $\theta$ is close to $1-\alpha$, for
any value of the parameters $(\gamma_5 - \gamma_4, \gamma_6 - \gamma_4)$  outside this square.

Our conclusion is that we may search for the minimum CP of the CI described in Section 2 as follows.
Let $min_1$ denote the estimate of the CP of this CI minimized over
$(\gamma_{4},\gamma_{5},\gamma_{6}) \in [-0.25, 0.25]^3$. Also, let $min_2$ denote the
estimate of the CP of this CI for $\gamma_4 = 1000$ (so that $\text{Pr}\big(F_{\tau} > \ell_{\tau} \big) \approx 1$)
and $(\gamma_5 - \gamma_4, \gamma_6 - \gamma_4) \in [-0.2,0.2]^2$. Then, our estimate of the minimum CP
of the CI described in Section 2 is the smaller of $min_1$ and $min_2$. Using this procedure,
with $M=10000$ simulation runs for each parameter value,
we found that $min_1 = 0.4385$ and $min_2 = 0.5175$, so that
the
minimum CP of the CI described in Section 2 is estimated to be 0.4385.
The minimum CP is achieved for $(\gamma_4,\gamma_5,\gamma_6) \in [-0.25,0.25]^3$.
A detailed description of this CP function for $(\gamma_4,\gamma_5,\gamma_6) \in [-0.25,0.25]^3$
is provided in Section 2.


\newpage

\begin{center}
\large{{\bf Appendix E: Derivation of convenient expressions for the conditional probabilities described in Section 4}}
\end{center}

As in Appendices B and C, let $\boldsymbol{Q}=\boldsymbol{\hat\tau}/\sigma$,
$D=m \hat\Sigma^2 / \sigma^2$. Also, let $\Phi$ denote the $N(0,1)$ distribution function.
The test statistics $F_{\tau}$ and $F_{\xi}$ are both functions of $(\boldsymbol{Q}, D)$.
In this appendix, we make this explicit by writing
$F_{\tau}=F_{\tau}(\boldsymbol{Q}, D)$ and $F_{\xi}=F_{\xi}(\boldsymbol{Q}, D)$.

The following are convenient expressions for the conditional probabilities $p_{\tau}$, $p_{\xi}$ and $p$ described in Section 4.
\begin{itemize}
\item
Let $e_{\tau}=t(m+k) \sqrt{\left(d + \boldsymbol{q}^{\top}\boldsymbol{V_{22}}^{-1}\boldsymbol{q}\right)/(m+k)} \sqrt{v^{*}}$. Note that
\begin{align*}
&p_{\tau}(\boldsymbol{q},d)= \\
&\Phi\bigg( \left(\boldsymbol{v_{21}}^{\top}\boldsymbol{V_{22}}^{-1}(\boldsymbol{\tau}/\sigma) \ + \ e_{\tau} \right) \Big / \sqrt{v^{*}} \bigg)
- \Phi\bigg( \left(\boldsymbol{v_{21}}^{\top}\boldsymbol{V_{22}}^{-1}(\boldsymbol{\tau}/\sigma) \ - \ e_{\tau} \right) \Big / \sqrt{v^{*}} \bigg)
\end{align*}
if $F_{\tau}(\boldsymbol{q}, d)\leq\ell_{\tau}$; otherwise $p_{\tau}(\boldsymbol{q},d)=0$.

\item
Let $e_{\xi}=t(m+k-1) \sqrt{\left(d + \boldsymbol{q}^{\top}\boldsymbol{U}^{\top}\boldsymbol{W_{22}}^{-1}\boldsymbol{U}\boldsymbol{q}\right)/(m+k-1)} \sqrt{w^{*}}$
and $\boldsymbol{s_{21}}=\boldsymbol{v_{21}}
- \boldsymbol{C_{\tau}}^{\top} (\boldsymbol{X}^{\top}\boldsymbol{X})^{-1} \boldsymbol{C_{\xi}} \boldsymbol{W_{22}}^{-1}
\boldsymbol{w_{21}}$.
Note that
\begin{align}
&p_{\xi}(\boldsymbol{q},d)= \notag\\
&\Phi\bigg( \left(\boldsymbol{w_{21}}^{\top}\boldsymbol{W_{22}}^{-1}(\boldsymbol{\xi}/\sigma) +  \boldsymbol{s_{21}}^{\top}\boldsymbol{V_{22}}^{-1}(\boldsymbol{\tau}/\sigma-\boldsymbol{q})
          \ + \ e_{\xi} \right) \Big/ \sqrt{w^{*}-\boldsymbol{s_{21}}^{\top}\boldsymbol{V_{22}}^{-1}\boldsymbol{s_{21}}} \bigg)  \notag \\
\label{prob_ksi}
&- \Phi\bigg( \left(\boldsymbol{w_{21}}^{\top}\boldsymbol{W_{22}}^{-1}(\boldsymbol{\xi}/\sigma) +  \boldsymbol{s_{21}}^{\top}\boldsymbol{V_{22}}^{-1}(\boldsymbol{\tau}/\sigma-\boldsymbol{q})
          \ - \ e_{\xi} \right) \Big/ \sqrt{w^{*}-\boldsymbol{s_{21}}^{\top}\boldsymbol{V_{22}}^{-1}\boldsymbol{s_{21}}} \bigg)
\end{align}
if $F_{\tau}(\boldsymbol{q}, d)>\ell_{\tau}$ and $F_{\xi}(\boldsymbol{q}, d)\leq\ell_{\xi}$; otherwise $p_{\xi}(\boldsymbol{q},d)=0$.

\item
Finally, let $e = t(m) \sqrt{d/m} \sqrt{v_{11}}$ and note that
\begin{align*}
&p(\boldsymbol{q},d)= \\
&\Phi\bigg( \left(\boldsymbol{v_{21}}^{\top}\boldsymbol{V_{22}}^{-1}(\boldsymbol{\tau}/\sigma-\boldsymbol{q}) \ + \ e \right) \Big / \sqrt{v^{*}} \bigg)
- \Phi\bigg( \left(\boldsymbol{v_{21}}^{\top}\boldsymbol{V_{22}}^{-1}(\boldsymbol{\tau}/\sigma-\boldsymbol{q}) \ - \ e \right) \Big / \sqrt{v^{*}} \bigg)
\end{align*}
if $F_{\tau}(\boldsymbol{q}, d)>\ell_{\tau}$ and $F_{\xi}(\boldsymbol{q}, d)>\ell_{\xi}$; otherwise $p(\boldsymbol{q},d)=0$.
\end{itemize}
\medskip

We now present the proof of the formula for $p_{\xi}(\boldsymbol{q},d)$.
The proofs of the formulas for $p_{\tau}(\boldsymbol{q},d)$ and $p(\boldsymbol{q},d)$ are similar, but simpler. For the sake of brevity, we omit these proofs.
We use the notation
\begin{equation*}
{\cal I}({\cal A}) =
\begin{cases}
1 &\text{if } {\cal A} \ \ \text{is true} \\
0 &\text{if } {\cal A} \ \ \text{is false}
\end{cases}
\end{equation*}
where ${\cal A}$ is an arbitrary statement. This is similar to the Iverson bracket notation (Knuth, 1992).
Observe that
\begin{align*}
&p_{\xi}(\boldsymbol{q},d)=\text{Pr} \left( \theta \in I_{\xi}, \, B \ \vert \ \boldsymbol{Q}=\boldsymbol{q}, \ D=d \right) \\
& \ \ \ \ \ \ \ \ \ \
=\text{Pr} \left( \theta \in I_{\xi} \ , \ F_{\tau}(\boldsymbol{Q}, D)>\ell_{\tau} \ , \ F_{\xi}(\boldsymbol{Q}, D)\leq \ell_{\xi}  \  \vert \ \boldsymbol{Q}=\boldsymbol{q}, \ D=d \right) \\
& \ \ \ \ \ \ \ \ \ \
=\text{E} \left( {\cal I} \left(\theta \in I_{\xi}\right)  \ {\cal I} \left(F_{\tau}(\boldsymbol{Q}, D)>\ell_{\tau} , F_{\xi}(\boldsymbol{Q}, D)\leq \ell_{\xi}\right)  \  \vert \ \boldsymbol{Q}=\boldsymbol{q}, \ D=d \right) \\
& \ \ \ \ \ \ \ \ \ \
=\text{E} \left( {\cal I} \left(\theta \in I_{\xi}\right)  \ {\cal I} \left(F_{\tau}(\boldsymbol{q}, d)>\ell_{\tau} , F_{\xi}(\boldsymbol{q}, d)\leq \ell_{\xi}\right)  \  \vert \ \boldsymbol{Q}=\boldsymbol{q}, \ D=d \right)
\end{align*}
by the substitution theorem for conditional expectations (see eg. p.9 of Bickel $\&$ Doksum, 1977). Thus
\begin{equation*}
p_{\xi}(\boldsymbol{q},d)=
\begin{cases}
\text{Pr} \left( \theta \in I_{\xi} \ \vert \ \boldsymbol{Q}=\boldsymbol{q}, \ D=d \right) \ \ \text{if} \ F_{\tau}(\boldsymbol{q}, d)>\ell_{\tau} \ \text{and} \ F_{\xi}(\boldsymbol{q}, d)\leq \ell_{\xi} \\
0 \ \ \text{otherwise.}
\end{cases}
\end{equation*}
It follows from \eqref{event_theta_in_I_ksi} that $\text{Pr} \left( \theta \in I_{\xi} \ \vert \ \boldsymbol{Q}=\boldsymbol{q}, \ D=d \right)$ is equal to
\begin{align}
&\text{Pr}\left( \boldsymbol{a}^{\top}\boldsymbol{\gamma} \in \left[ \boldsymbol{a}^{\top}\boldsymbol{G_{\xi}\hat\gamma} \ \pm \
t(m+k-1) \sqrt{\frac{D + \boldsymbol{Q}^{\top}\boldsymbol{U}^{\top}\boldsymbol{W_{22}}^{-1}\boldsymbol{U}\boldsymbol{Q}}{m+k-1}} \sqrt{w^{*}} \right]
\Bigg\vert \ \boldsymbol{Q}=\boldsymbol{q}, \ D=d \right) \notag \\
\label{prob_ksi_1}
&= \text{Pr}\left( \boldsymbol{a}^{\top}\boldsymbol{\gamma} \in \left[ \boldsymbol{a}^{\top}\boldsymbol{G_{\xi}\hat\gamma} \ \pm \
t(m+k-1) \sqrt{\frac{d + \boldsymbol{q}^{\top}\boldsymbol{U}^{\top}\boldsymbol{W_{22}}^{-1}\boldsymbol{U}\boldsymbol{q}}{m+k-1}} \sqrt{w^{*}} \right]
\Bigg\vert \ \boldsymbol{Q}=\boldsymbol{q}, \ D=d \right) \notag \\
\end{align}
by the substitution theorem for conditional expectations. Since $\boldsymbol{\hat\gamma}$ and $D$ are independent random vectors, \eqref{prob_ksi_1} is equal to
\begin{align}
&\text{Pr}\left( \boldsymbol{a}^{\top}\boldsymbol{\gamma} \in \left[ \boldsymbol{a}^{\top}\boldsymbol{G_{\xi}\hat\gamma} \ \pm \
t(m+k-1) \sqrt{\frac{d + \boldsymbol{q}^{\top}\boldsymbol{U}^{\top}\boldsymbol{W_{22}}^{-1}\boldsymbol{U}\boldsymbol{q}}{m+k-1}} \sqrt{w^{*}} \right]
\ \Bigg\vert \ \boldsymbol{Q}=\boldsymbol{q} \right) \notag \\
&= \text{Pr}\left( \boldsymbol{a}^{\top}\boldsymbol{\gamma} \in \left[ \boldsymbol{a}^{\top}\boldsymbol{G_{\xi}\hat\gamma} \ \pm \ e_{\xi} \right]
\ \vert \ \boldsymbol{Q}=\boldsymbol{q} \right) \notag \\
\label{prob_ksi_2}
&= \text{Pr}\left( \boldsymbol{a}^{\top}\boldsymbol{\gamma}- e_{\xi} \leq \boldsymbol{a}^{\top}\boldsymbol{G_{\xi}\hat\gamma} \leq \boldsymbol{a}^{\top}\boldsymbol{\gamma}+ e_{\xi}
\ \vert \ \boldsymbol{Q}=\boldsymbol{q} \right)
\end{align}
Note that the random vectors $\boldsymbol{a}^{\top}\boldsymbol{G_{\xi}\hat\gamma}$ and $\boldsymbol{Q}$ have the following multivariate normal distribution.
\begin{equation*}
\begin{bmatrix}  \boldsymbol{a}^{\top}\boldsymbol{G_{\xi}\hat\gamma} \\ \boldsymbol{Q} \end{bmatrix} \sim
N \left( \begin{bmatrix} \boldsymbol{a}^{\top}\boldsymbol{\gamma}-\boldsymbol{w_{21}}^{\top}\boldsymbol{W_{22}}^{-1}(\boldsymbol{\xi}/\sigma) \\
\boldsymbol{\tau} / \sigma \end{bmatrix},
\begin{bmatrix} w^{*} & \boldsymbol{s_{21}}^{\top} \\ \boldsymbol{s_{21}} & \boldsymbol{V_{22}} \end{bmatrix}  \right)
\end{equation*}
Thus, the distribution of $\boldsymbol{a}^{\top}\boldsymbol{G_{\xi}\hat\gamma}$ conditional on $\boldsymbol{Q}=\boldsymbol{q}$ is
\begin{equation*}
N \Big( \boldsymbol{a}^{\top}\boldsymbol{\gamma}-
\boldsymbol{w_{21}}^{\top}\boldsymbol{W_{22}}^{-1}(\boldsymbol{\xi}/\sigma) -
\boldsymbol{s_{21}}^{\top}\boldsymbol{V_{22}}^{-1}(\boldsymbol{\tau}/\sigma - \boldsymbol{q}) \ , \ w^{*}-\boldsymbol{s_{21}}^{\top}\boldsymbol{V_{22}}^{-1}\boldsymbol{s_{21}} \Big)
\end{equation*}
Hence, \eqref{prob_ksi_2} is equal to
\begin{align*}
\begin{split}
&\text{Pr}\left( \frac{\boldsymbol{a}^{\top}\boldsymbol{\gamma}- e_{\xi}-\text{E}(\boldsymbol{a}^{\top}\boldsymbol{G_{\xi}\hat\gamma} \ \vert \ \boldsymbol{Q}=\boldsymbol{q})}
{\sqrt{\text{Var}(\boldsymbol{a}^{\top}\boldsymbol{G_{\xi}\hat\gamma} \ \vert \ \boldsymbol{Q}=\boldsymbol{q})}}
\leq
\frac{\boldsymbol{a}^{\top}\boldsymbol{G_{\xi}\hat\gamma} -\text{E}(\boldsymbol{a}^{\top}\boldsymbol{G_{\xi}\hat\gamma} \vert \ \boldsymbol{Q}=\boldsymbol{q})}
{\sqrt{\text{Var}(\boldsymbol{a}^{\top}\boldsymbol{G_{\xi}\hat\gamma} \vert \ \boldsymbol{Q}=\boldsymbol{q})}} \right.\\
& \ \ \ \ \ \ \ \ \ \ \ \ \ \ \ \ \left. \leq
\frac{\boldsymbol{a}^{\top}\boldsymbol{\gamma}+e_{\xi}-\text{E}(\boldsymbol{a}^{\top}\boldsymbol{G_{\xi}\hat\gamma} \vert \ \boldsymbol{Q}=\boldsymbol{q})}
{\sqrt{\text{Var}(\boldsymbol{a}^{\top}\boldsymbol{G_{\xi}\hat\gamma} \vert \ \boldsymbol{Q}=\boldsymbol{q})}} \bigg\vert \ \boldsymbol{Q}=\boldsymbol{q}\right)
\end{split} \\
\begin{split}
& = \text{Pr}\left( \frac{\boldsymbol{w_{21}}^{\top}\boldsymbol{W_{22}}^{-1}(\boldsymbol{\xi}/\sigma)
+  \boldsymbol{s_{21}}^{\top}\boldsymbol{V_{22}}^{-1}(\boldsymbol{\tau}/\sigma-\boldsymbol{q}) \ - \ e_{\xi}}
{\sqrt{w^{*}-\boldsymbol{s_{21}}^{\top}\boldsymbol{V_{22}}^{-1}\boldsymbol{s_{21}}}}
\leq
Z
\right.\\ & \ \ \ \ \ \ \ \ \ \ \ \ \ \ \ \left.
\leq
\frac{\boldsymbol{w_{21}}^{\top}\boldsymbol{W_{22}}^{-1}(\boldsymbol{\xi}/\sigma)
+  \boldsymbol{s_{21}}^{\top}\boldsymbol{V_{22}}^{-1}(\boldsymbol{\tau}/\sigma-\boldsymbol{q}) \ + \ e_{\xi}}
{\sqrt{w^{*}-\boldsymbol{s_{21}}^{\top}\boldsymbol{V_{22}}^{-1}\boldsymbol{s_{21}}}} \right), \ \text{where} \ Z \sim N(0,1),
\end{split}
\end{align*}
which is equal to \eqref{prob_ksi}.

\bigskip

\begin{center}
\sl{References}
\end{center}

\smallskip

\rf BICKEL, P.J. \& DOKSUM, K.A. (1977). Mathematical Statistics. Holden-Day, Oakland, California.

\smallskip

\rf GIRI, K. \& KABAILA, P. (2008). The coverage probability of confidence intervals
in $2^r$ factorial experiments after a preliminary hypothesis test. {\sl Australian \& New
Zealand Journal of Statistics} {\bf 50}, 69--79.

\smallskip

\rf KABAILA, P. (2009). The coverage properties of confidence regions after model selection.
{\sl International Statistical Review} {\bf 77}, 405--414.

\smallskip

\rf KABAILA, P. \& FARCHIONE, D. (2012).
The minimum coverage probability of confidence intervals in regression
after a preliminary F test.
{\sl Journal of Statistical Planning and Inference} {\bf 142}, 956--964.

\smallskip

\rf KNUTH, D.E. (1992). Two notes on notation. {\sl American Mathematical Monthly}
{\bf 99}, 403--422

\smallskip

\rf MILLIKEN, G.A. \& JOHNSON, D.E. (2002).
{\sl Analysis of Messy Data. Volume III: Analysis of Covariance}.
Chapman \& Hall/CRC, Boca Raton, Florida.

\smallskip

\rf ROSS, S.M. (2000). {\sl Introduction to Probability Models, Seventh Edition}.
Harcourt/Academic Press, San Diego, California.

\end{document}